\documentclass[letterpaper]{article} % DO NOT CHANGE THIS
\usepackage{aaai2026}  % DO NOT CHANGE THIS
\usepackage{times}  % DO NOT CHANGE THIS
\usepackage{helvet}  % DO NOT CHANGE THIS
\usepackage{inconsolata}  % monospace font (Consolas-like)
\usepackage[hyphens]{url}  % DO NOT CHANGE THIS
\usepackage{graphicx} % DO NOT CHANGE THIS
\usepackage{natbib}  % DO NOT CHANGE THIS AND DO NOT ADD ANY OPTIONS TO IT
\usepackage{caption} % DO NOT CHANGE THIS AND DO NOT ADD ANY OPTIONS TO IT
\usepackage{algorithm}
\usepackage{algorithmic}
\usepackage{soul}  % provides \ul{} for punchline underlining (not on AAAI forbidden list)
\usepackage{booktabs}  % provides \toprule, \midrule, \bottomrule for tables

\usepackage{newfloat}
\usepackage{listings}
\DeclareCaptionStyle{ruled}{labelfont=normalfont,labelsep=colon,strut=off} % DO NOT CHANGE THIS
\floatstyle{ruled}
\newfloat{listing}{tb}{lst}{}
\floatname{listing}{Listing}

\title{Co-Skill: A Collaborative Communication Framework for Skill Evolution}

\author{
    Yilin Ma*$^{1}$†, Yanqi Pan*$^{1}$†, Weihao Yang*$^{1}$†, Peixin Zeng$^{1}$, Jiannan Xu$^{1}$, Hao Huang$^{1}$, Wen Xia$^{1}$‡
}
\affiliations{
    $^{1}$Harbin Institute of Technology, Shenzhen
    
    †: \texttt{yilin0552@gmail.com,deadpooldeathmine@gmail.com,weihao.yang00@hotmail.com}
    ‡: \texttt{xiawen@hit.edu.cn}
}

\usepackage{bibentry}
\begin{document}

\maketitle

\begin{abstract}
Agent evolution through skills becomes critical for LLM-based agents to iteratively improve task success rate. Hybrid evolution is a cost-efficient paradigm where a cloud LLM analyzes and generates skills while an edge SLM executes and internalizes them. However, existing hybrid methods, such as SkillRL, still suffer from low success rate and high token usage. We find this stems from \emph{blind communication}: the cloud cannot perceive the edge's execution capability, while the edge does not understand the cloud's analysis needs. 

We thus propose the \emph{Collaborative Communication Framework} (CCF) to achieve effective edge-cloud evolution. CCF is realized via three techniques: (1) a \emph{cloud-aware prefix-merged trajectory trie} where the edge compresses trajectories by merging shared prefixes and pinpointing divergence points for efficient cloud analysis, (2) an \emph{edge-aware progressive skill tree} where the cloud progressively builds a hierarchical skill tree to match edge SLM execution capability, and (3) a \emph{collaborative skill evolution} scheme upon these two trees that evolves cloud LLM and edge SLM in a separated way to jointly improve task success rate. Experiments across ALFWorld and WebShop show that CCF reduces LLM+SLM tokens by 15.6\%--41.9\% over state-of-the-art hybrid methods while consistently improving 25.8\%--76.4\% task success rate.
\end{abstract}


\section{Introduction}

Agent evolution, the iterative improvement of capabilities through accumulated experience, is increasingly critical for LLM-based agents.
Evolution operates via \textit{skills}: structured, reusable knowledge that agents acquire and refine across tasks~\cite{What-are-skills}.
Through skill evolution, each cycle improves task success rate and reduces token consumption, making it the central mechanism for building cost-efficient agents~\cite{xskill,arise,Memento-skill}.

Existing agent evolution follows three paradigms.
First, \textit{cloud-only evolution} runs all skill learning and inference on frontier, cloud-based LLMs (e.g., DeepSeek-v4)~\cite{ni2026trace2skilldistilltrajectorylocallessons}, achieving strong performance but at high cost.
Second, \textit{edge-only evolution} fine-tunes a small language model (SLM) locally via Reinforcement Learning (RL)~\cite{lu2026skill0incontextagenticreinforcement}. It is fast and cheap, yet generally fails on complex tasks due to limited reasoning capacity. Recently, \textit{hybrid evolution}, such as SkillRL~\cite{xia2026skillrlevolvingagentsrecursive}, combines the best of the above two paradigms: it uses a cloud LLM to analyze and generate skills, while using an edge SLM to execute and internalize generated skills via RL. Consequently, \ul{hybrid evolution is a promising cost-efficient paradigm}~\cite{xia2026skillrlevolvingagentsrecursive, The-Internet-of-Agentic-AI} \ul{and the focus of our work}.

However, existing hybrid evolution suffers from two challenges.
The first is \textit{cloud communication redundancy}: the edge agent uploads raw trajectories for cloud analysis, yet over 25.0\%--41.8\% of uploaded tokens repeat shared trajectory prefixes, wasting cloud tokens and increasing latency.
The second is \textit{edge execution deficiency}: the cloud LLM generates skills based on its own reasoning, without perceiving what the edge can actually execute. The edge receives undifferentiated instructions, leading to brittle execution.

We find that both challenges stem from a single root cause: \textbf{blind communication}.
On the one hand, the edge is blind to the cloud's analysis needs; it dumps raw trajectories undifferentiated, forcing the cloud to sift redundant context.
On the other hand, the cloud is blind to the edge's execution capability; it sends guidance without knowing what the edge can execute.
\ul{Our key insight is a \emph{Collaborative Communication Framework (CCF)} that replaces blind communication with mutual awareness,} where each side perceives the other's capability and adapts its communication accordingly.

Specifically, CCF adopts three techniques.
First, a \textit{cloud-aware prefix-merged trajectory trie}, where the edge proactively merges shared trajectory prefixes and surfaces divergence points as tree branches, eliminating redundant upload while highlighting the key differences (among trajectories), thereby addressing cloud communication redundancy.
Second, an \textit{edge-aware progressive skill tree}, where the cloud progressively generates skills in a tree manner: It generates skill context from the root, high-level goals to low-level, concrete task-specific guidance, so the cloud LLM can determine the edge execution ability progressively (e.g., when an edge agent can achieve a high success rate), thereby addressing edge execution deficiency. Finally, CCF adopts \textit{collaborative skill evolution} atop these two trees to jointly evolve cloud and edge LLMs. Specifically, the cloud maintains a hierarchical skill library, 
while the edge learns improved skills with an optional RL module to internalize the evolution. 
% while the edge performs RL to internalize stabilized skills and refine skills to match the ability. 

We incarnate the idea of CCF via \textsc{Co-Skill}. Experiments across ALFWorld and WebShop show that \textsc{Co-Skill} reduces SLM+LLM tokens by 15.6\%--41.9\% compared to state-of-the-art hybrid agents (e.g., SkillRL) while improving 25.8\%--76.4\% task success rates.

This paper makes the following contributions:

\begin{itemize}
\item We identify that \emph{blind communication} is the root cause of the deficiency of existing hybrid evolution methods.

\item We propose the Collaborative Communication Framework (CCF) for bidirectional cloud-edge evolution.
\item We design \textsc{Co-Skill} to realize CCF with a \textit{cloud-aware prefix-merged trajectory trie}, an \textit{edge-aware progressive skill tree}, and \textit{collaborative skill evolution}.
\item Experiments on ALFWorld and WebShop show that \textsc{Co-Skill} reduces SLM+LLM token usage while improving task success rate compared to cloud-edge baselines.
\end{itemize}

\section{Related Work and Motivation}

\begin{figure}[t]
\centering
\includegraphics[width=\linewidth]{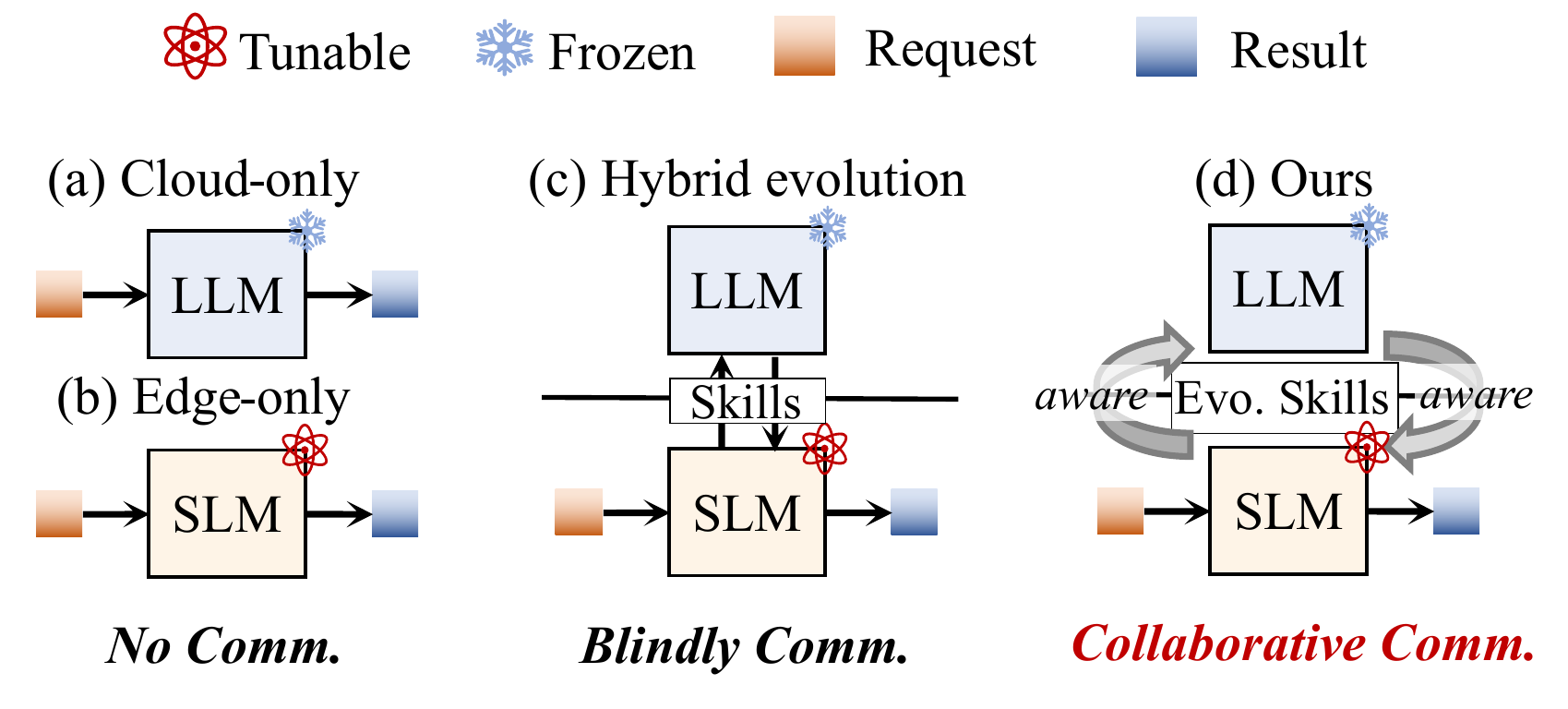}
\caption{The evolution paradigm comparison. The cloud-only/edge-only paradigm hardly balances evolution and token consumption~\cite{lu2026skill0incontextagenticreinforcement}, while the state-of-the-art hybrid evolution~\cite{xia2026skillrlevolvingagentsrecursive} achieves the suboptimal balance due to blind communication between the edge SLM and cloud LLM. Our proposed CCF achieves the best results through edge-cloud collaborative communication.}
\label{fig:arch_compare}

\end{figure}

In this section, we compare agent skill evolution paradigms (Figure~\ref{fig:arch_compare}), and we describe them below.

\paragraph{Cloud-Based LLM and Evolution.}

Frontier, cloud-based LLMs, such as DeepSeek-v4~\cite{deepseekai2026deepseekv4highlyefficientmilliontoken}, power agent reasoning, from task decomposition to skill generation.
Inference engines such as vLLM~\cite{kwon2023efficient} and SGLang~\cite{zheng2024sglang} use prefix caching to reuse KV-cache across identical prefixes.
This mechanism works well for multi-turn conversations with an incremental prefix, but is under-exploited when analyzing different trajectories whose prefixes are not naturally aligned.

Recently, cloud LLMs have also been used to evolve agent skills.
Trace2Skill~\cite{ni2026trace2skilldistilltrajectorylocallessons} distills skills from execution trajectories, accumulating reusable knowledge across episodes. However, trajectories are processed blindly, which may violate the caching mechanism of the inference engine. Moreover, cloud LLMs are used for both analysis and execution, thereby increasing the evolution cost (especially under such multi-round trajectory analysis).

% However, shared prefixes across trajectories are . Each trajectory is sent independently, forcing the cloud to repeatedly process overlapping context and wasting bandwidth. 

\paragraph{Edge-Based Agent and Evolution.}

Edge-based agents run small language models (SLM) locally, with a typical size ranging 3B--30B~\cite{qwen3technicalreport,lu2026skill0incontextagenticreinforcement}.
They typically follow the ReAct~\cite{yao2023react} paradigm, interleaving reasoning with tool invocation for multi-step tasks.
The lightweight deployment offers low latency, but limited capacity constrains reasoning.

To improve edge SLMs, recent work evolves them by reinforcement learning (RL). Skill0~\cite{lu2026skill0incontextagenticreinforcement} applies GRPO-style RL to train edge agents without cloud dependency. Skill discovery methods~\cite{eyal2019dads,gregor2016variational} learn reusable behaviors, and RL with verifiable rewards~\cite{da2025agent} enables outcome-driven fine-tuning. However, the edge SLMs usually have poor reasoning and execution capabilities due to their limited parameter constraints. As a result, even with RL, they still suffer from a low task success rate.

\begin{figure*}[t]
\centering
\includegraphics[width=\linewidth]{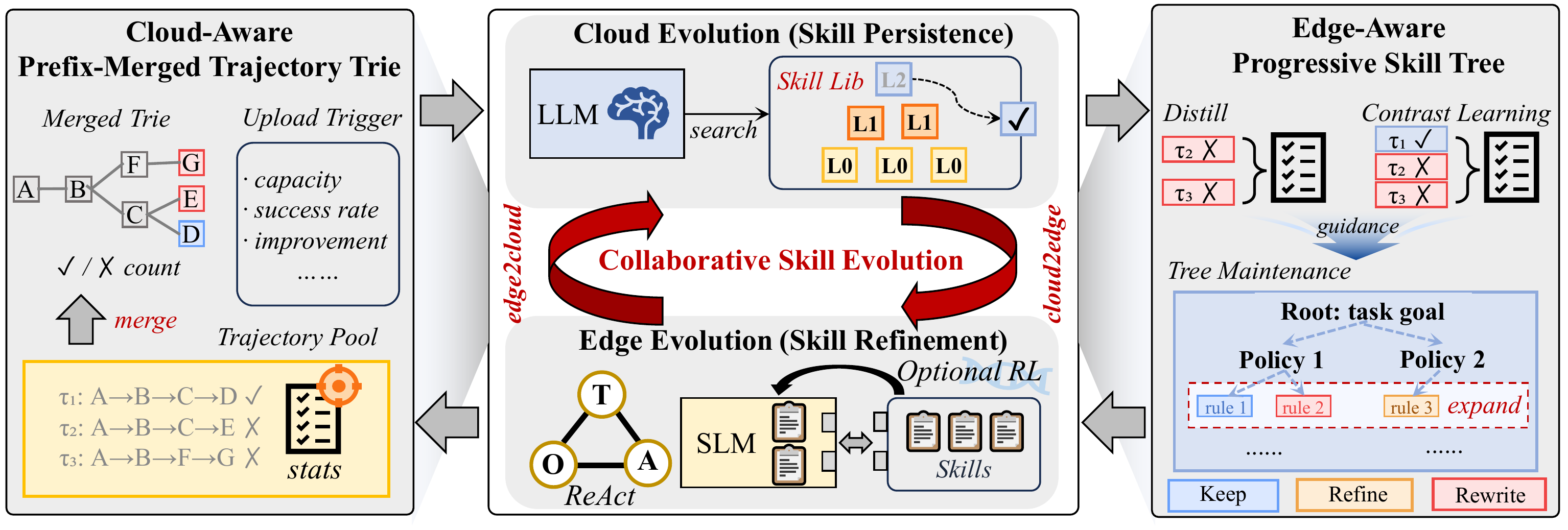}
\caption{Overview of the \textsc{Co-Skill} architecture. \textsc{Co-Skill} includes three major components: a Cloud-aware prefix-merged trajectory trie that compresses edge-to-cloud traffic; an Edge-aware progressive skill tree that organizes cloud-to-edge skills; and a collaborative skill evolution method that jointly evolves edge-cloud models/agents.}
\label{fig:overview}
\end{figure*}
% these skills are typically generated by LLMs, and their granularity often mismatches the edge's reason and execution capability. Consequently, the edge SLMs/agents often fail to interpret the skills. Even with RL, they still suffer from low success rate.

\paragraph{Hybrid Evolution.}
To address the above deficiencies, hybrid evolution aims to combine cloud LLM analysis with edge SLM execution~\cite{xia2026skillrlevolvingagentsrecursive}. A typical workflow proceeds as: (1) the cloud LLM initializes skills for the edge SLM, (2) the edge SLM/agent executes tasks using those skills, (3) execution traces are uploaded to the cloud, (4) the cloud analyzes traces and identifies failures, (5) the cloud generates refined skills and sends them back, (6) the edge may perform RL to absorb updated skills. However, we find that blindly combining these two evolution mechanisms does not necessarily improve the task success rate (as shown in Figure~\ref{fig:main_results}), nor does it reduce the usage of SLM tokens (as shown in Figure~\ref{tab:token_cost}), thus remaining a large room to improve. 

The challenges lie in two aspects: (1) \emph{Cloud communication redundancy}. Trajectories sharing similar prefixes are uploaded to the cloud LLM, which redundantly reasons over overlapping content and infers divergence points itself, wasting tokens and reducing efficiency. Our observation shows that this can cause 15.3\%--40.5\% wasted tokens (Figure~\ref{fig:compression}). (2) \emph{Edge execution efficiency}. By applying LLM-generated skills, edge SLM either fails to execute skills (i.e., low success rate) or spends more tokens to reason the given skills (i.e., 33.1\%--106.1\% more tokens compared to ours on ALFWorld, Figure~\ref{tab:token_cost}). This is primarily because the misalignment between LLM-generated skills and the SLM's ability.

\paragraph{Root Cause.} We find these challenges are rooted in \emph{blind communication}, where the edge agent fails to provide LLM-friendly trajectories (e.g., with branches highlighted), thus wasting its tokens and reasoning ability, while the cloud LLM fails to be aware of SLM's reasoning and execution abilities, thus generating SLM-unfriendly skills that hinder the accuracy and decrease edge SLM execution efficiency.

% to generate skills 

% cannot see what the cloud needs to analyze efficiently

% the communication between cloud and edge remains a blind pipe: the edge cannot see what the cloud needs to analyze efficiently, and the cloud cannot perceive what the edge is capable of executing. Neither side tracks what the other has already learned.

\section{Method}

We thereby present \textsc{Co-Skill} to achieve accurate and cost-efficient hybrid agent. The core of \textsc{Co-Skill} is \emph{Collaborative Communication Framework (CCF)}, which coordinates edge and cloud models through bidirectional awareness.

\subsection{Overview: Collaborative Communication Framework (CCF)}

\paragraph{Components.} As Figure~\ref{fig:overview} shows, CCF consists of three components, each addressing one aspect of blind communication.
First, a \textit{cloud-aware prefix-merged trajectory trie} structures what the edge uploads, enabling the cloud to reason over key decision forks rather than sift through raw trajectories.
Second, an \textit{edge-aware progressive skill tree} structures what the cloud sends down, progressively generates skills to match edge SLM's execution capability.
Third, \textit{collaborative skill evolution} jointly evolves both sides: the cloud enriches the skill context without modifying its own parameters, while the edge continuously enhances its capabilities through updated skills, which is also allowed for an optional RL module to internalize the evolution.
% uses RL to internize skills and refine skills to match its ability.

\paragraph{Workflow.}
We then describe the workflow of \textsc{Co-Skill} with CCF's components. First, guided by the \textit{edge-aware progressive skill tree}, the edge model executes tasks and accumulates trajectories in a local pool.
Second, the \textit{cloud-aware prefix-merged trajectory trie} takes these trajectories, merges shared prefixes and surfaces divergence points, then uploads the compressed result to the cloud.
Third, the cloud analyzes failures at the trie forks and updates the skill tree by keeping, refining or adding nodes, then sends the revised tree back to the edge.
Finally, \textit{collaborative skill evolution} takes over: the cloud stores refined skill contexts, while the edge learns improved skills (and an optional RL, if required)

\paragraph{Advantages.}
As a result, \textsc{Co-Skill} with CCF addresses blind communication.
First, the cloud-aware trie addresses the first challenge (i.e., cloud communication redundancy), since it structures edge uploads so the cloud reasons over decision forks instead of sifting through raw trajectories.
Second, the edge-aware skill tree addresses the second challenge (i.e., edge execution deficiency), since it progressively generates skills to match edge SLM's execution capability. Finally, collaborative skill evolution lets the cloud refine the skill context and the edge absorb skills via RL, each operating where it excels, jointly optimizing execution efficiency.

\begin{figure}[t]
\centering
\includegraphics[width=\linewidth]{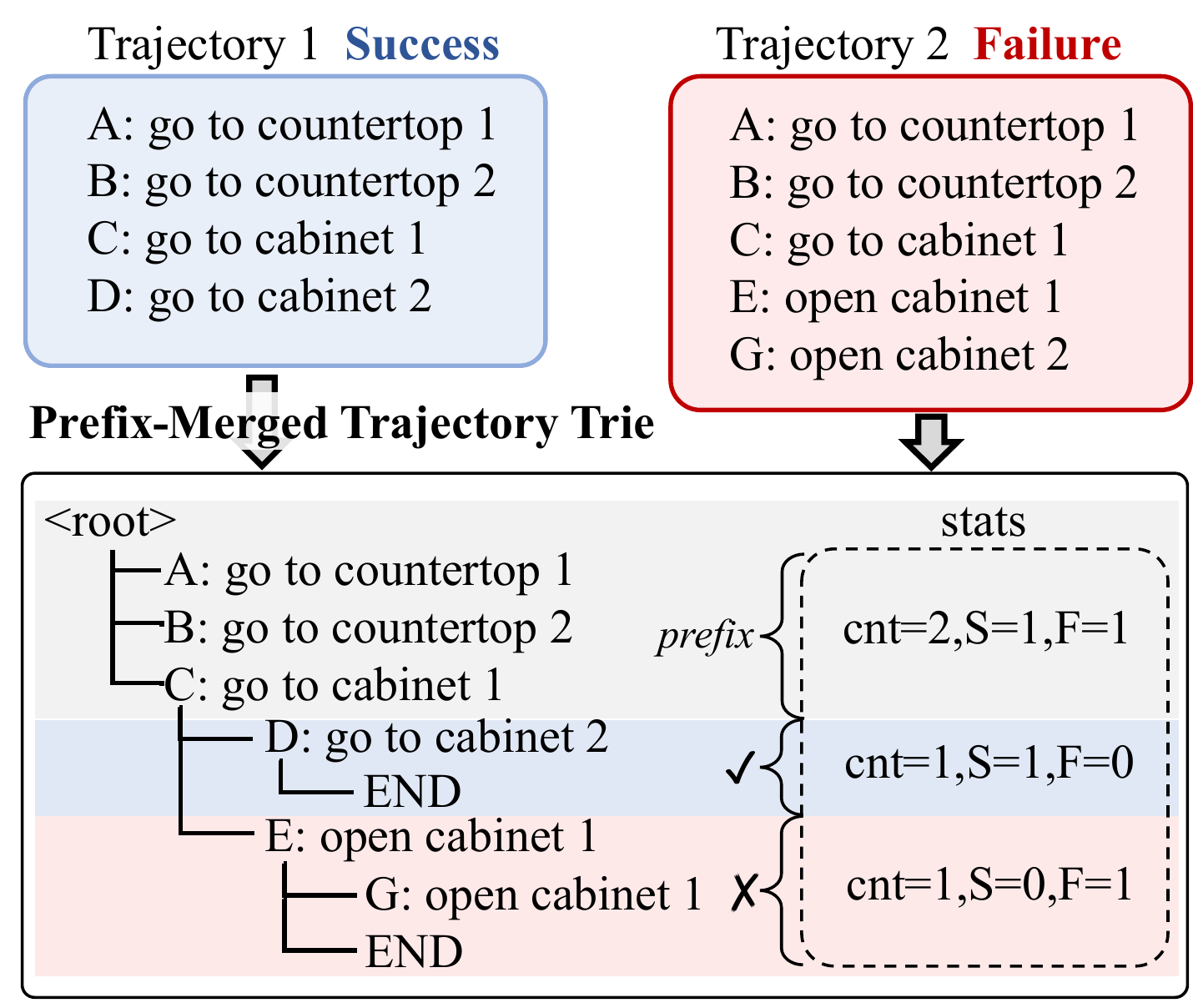}
\caption{Cloud-aware prefix-merged trie construction. Top: two raw trajectories with shared prefixes (A--C). Bottom: the trie collapses consensus and surfaces divergence at E. Annotations show [count, successes (S), failures (F)].}
\label{fig:trie_example}
\end{figure}
\subsection{Cloud-Aware Prefix-Merged Trajectory Trie}\label{sec:cloud-aware-trie}

The edge-to-cloud channel is blind to what the cloud needs to analyze. The edge uploads raw, plain trajectories without preprocessing (e.g., merging), thereby wasting up to 40.5\% of tokens.
We address this with a \textit{cloud-aware prefix-merged trajectory trie}, where the edge agent compresses repeated successes and surfaces divergence points, enabling the cloud to focus inference on the decision forks that matter.

\paragraph{Trajectory Trie Construction.}
The trie merges $N$ trajectories of the same task type through two operations. First, identical actions at the same state merge into a consensus node. Second, diverging actions branch into decision forks. Each node records the action, outcome counts, and success/failure distribution. Observation delta encoding further reduces tokens by transmitting only the text difference when it is shorter than half the full observation length.

Figure~\ref{fig:trie_example} illustrates this with two trajectories. First, they both share actions A, B, and C, which the trie collapses into a single consensus path. Second, at action C they diverge: Trajectory 1 proceeds to D and succeeds, while Trajectory 2 branches to E and G and fails. Third, the trie surfaces this divergence at E, with outcome counts directly visible. As a result, the cloud LLM immediately sees where the edge deviated without scanning two raw trajectories.

\paragraph{Trajectory Pool Merge.}
The pool manages trajectory accumulation on the edge. As shown in Figure~\ref{fig:trie_example}, trajectories are merged before upload. An upload is triggered when any of the following conditions holds. (1) A task type has accumulated at least 16 recent samples and its recent failure rate is at least 0.6. (2) Its recent success rate drops substantially or remains persistently below near-perfect performance. (3) The cumulative compressed payload reaches 50,000 tokens. All uploads are asynchronous.

\paragraph{Advantages.}
The trie is Cloud-Aware in two ways. First, it compresses consensus: shared paths are merged so the cloud never re-processes what the edge already masters. Second, it surfaces divergence: decision forks automatically mark where the cloud should focus its reasoning.
\ul{This turns a blind raw dump into a structured input that directs cloud inference to where it matters.}
As a result, the cloud no longer sifts through redundant trajectories but reasons directly over the decision forks that require attention.

\subsection{Edge-Aware Progressive Skill Tree}\label{sec:skill-tree}

The cloud-to-edge channel is blind to the edge's execution capability. In prior work~\cite{xia2026skillrlevolvingagentsrecursive}, the   cloud generates new strategies based on error reports from the edge. To prevent recurring failures, the cloud may produce increasingly complex descriptions. However, the edge model, with its limited capacity, often fails to parse these complex instructions and continues to execute incorrectly.
We address this with an \textit{edge-aware progressive skill tree}, where the cloud progressively generates skills at increasing depth, each level matching the edge's execution capability.

\paragraph{Overview.}
The cloud LLM builds the skill tree progressively in two stages. First, it generates a high-level description of the task from a few seed trajectories and sends this initial guidance to the edge. The edge executes with this guidance, and its trajectories flow through the trajectory pool and cloud-aware trie (\S\ref{sec:cloud-aware-trie}) back to the cloud. Second, the cloud inspects the trie to assess the edge's performance at each node: if accuracy is already sufficient, that branch stops growing. Only when accuracy falls short does the cloud expand the node into finer sub-steps, repeating this check at each new level. As a result, the tree grows only where the edge genuinely needs help, staying compact and matched to the edge's capability at every branch.

\begin{figure}[t]
\centering
\includegraphics[width=\linewidth]{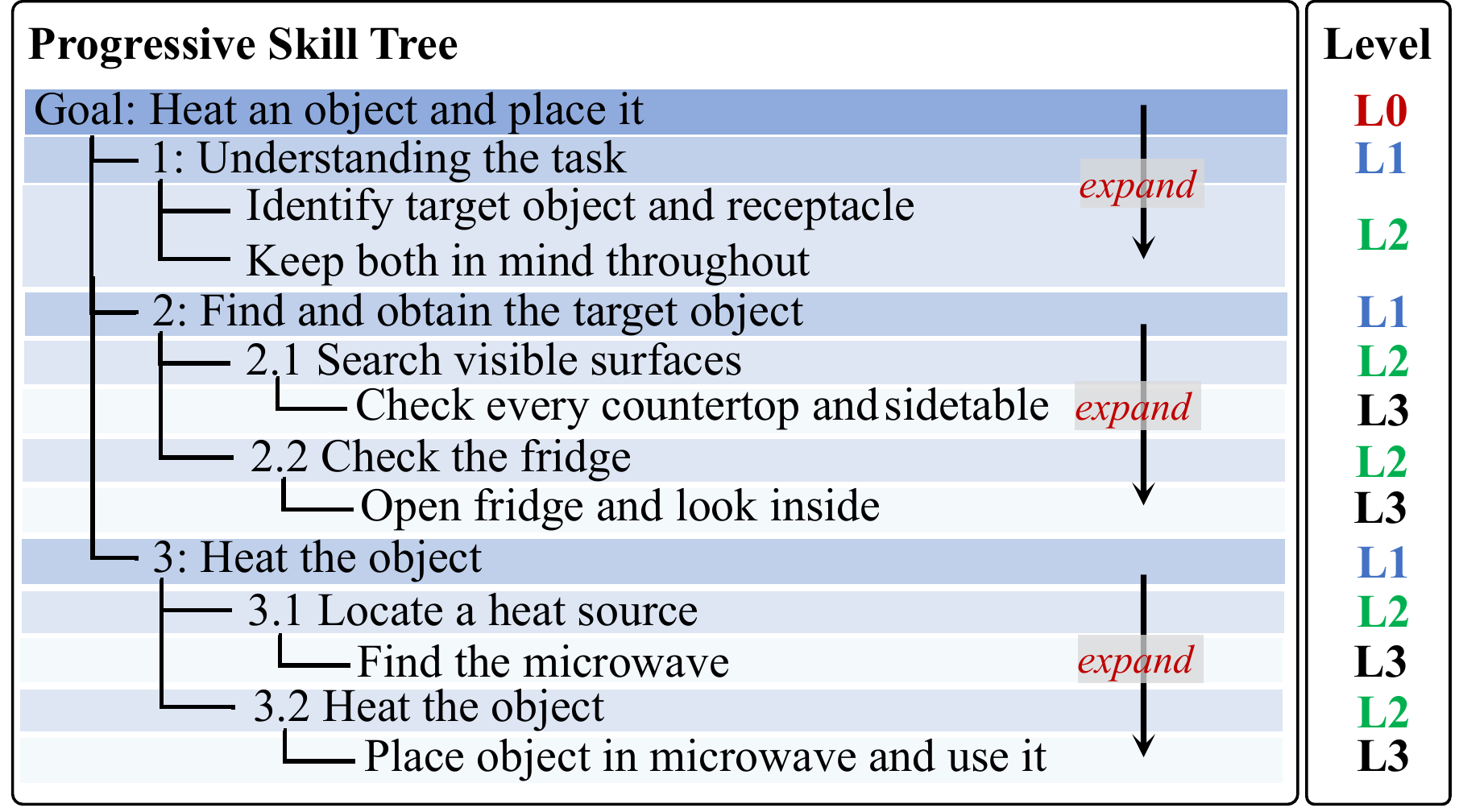}
\caption{Edge-aware progressive skill tree (an ALFWorld example). The cloud LLM progressively generates the skill by level (e.g., L0--L3), enabling it to progressively and precisely match the edge SLM's execution capability.}
\label{fig:skill_tree}
\end{figure}

Figure~\ref{fig:skill_tree} shows an example for heating an object in ALFWorld. The root level (L0) states the goal. Level 1 describes task understanding. Level 2 decomposes object retrieval into searching surfaces then checking the fridge. Level 3 covers heating via microwave. Note that this is the final skill tree, which means the edge SLM fails to achieve a promising success rate until cloud expands the tree to Level 3. We comprehensively study the tree depth and success rate in \S\ref{sec:ablation}.

% The edge starts at the root and descends only as deep as needed.

% \begin{figure}[t]
% \centering
% \begin{minipage}{1\linewidth}
% \centering{\small\textbf{Edge-Aware Progressive Skill Tree}}\vspace{-0.5em}
% \rule{\linewidth}{0.4pt}\vspace{0.2em}
% \begin{lstlisting}[
%     basicstyle=\footnotesize\ttfamily,
%     numbers=none,
%     frame=none,
%     xleftmargin=0pt,
%     aboveskip=0pt,
%     belowskip=0pt,
%     lineskip=0pt,
%     escapeinside={(@}{@)},
% ]
% Goal: Heat an object and place it (@\textbf{(L0)}@)
% |-- 1: Understand the task (@\textbf{(L1)}@)
% |   |-- Identify target object and receptacle
% |   |-- Keep both in mind throughout
% |-- 2: Find and obtain the target object (@\textbf{(L1)}@)
% |   |-- 2.1 Search visible surfaces (@\textbf{(L2)}@)
% |   |   |-- Check every countertop and sidetable (@\textbf{(L3)}@)
% |   |-- 2.2 Check the fridge (@\textbf{(L2)}@)
% |       |-- Open fridge and look inside (@\textbf{(L3)}@)
% |-- 3: Heat the object (@\textbf{(L1)}@)
%     |-- 3.1 Locate a heat source (@\textbf{(L2)}@)
%     |   |-- Find the microwave (@\textbf{(L3)}@)
%     |-- 3.2 Heat the object (@\textbf{(L2)}@)
%         |-- Place object in microwave and use it (@\textbf{(L3)}@)
% \end{lstlisting}
% \vspace{-0.5em}
% \rule{\linewidth}{0.4pt}
% \end{minipage}
% \caption{\textbf{Edge-aware progressive skill tree for an ALFWorld task.} \small The tree organizes skills by depth (L0--L3), enabling the edge to navigate at the appropriate granularity.}
% \label{fig:skill_tree}
% \end{figure}

\paragraph{Design Details.}
The cloud updates the tree after receiving the trie through three steps.
(1) \emph{Distill.} For each failure path, the cloud extracts a structured diagnosis: failure type, root cause, evidence step, corrective rule, and patch location. The patch location specifies the exact node where the correction belongs.
(2) \emph{Contrastive learning.} At each fork, the cloud compares successful and failed paths sharing a prefix to identify the behavioral difference that caused the divergence.
(3) \emph{Tree maintenance.} The cloud LLM decides the operation for each affected node: keep, refine with a targeted edit, or rewrite the subtree. The tree grows one level at a time, creating new children only when repeated failures with different root causes map to the same parent.

\begin{figure*}[t]
    \centering
    \includegraphics[width=\textwidth]{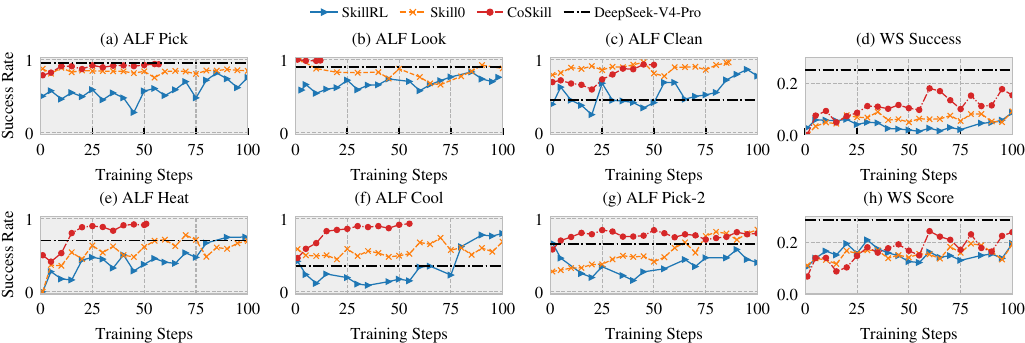}
    \caption{Task success rate across ALFWorld and WebShop workloads. We report success rates for the six ALFWorld task types (i.e., with ALF prefixes), together with the success rate and task score on WebShop (i.e., with WS prefixes).}
    \label{fig:main_results}
\end{figure*}

\paragraph{Why Tree Structure?}
An alternative approach is to let the cloud directly edit a flat, plain context document, as in prior work~\cite{xia2026skillrlevolvingagentsrecursive}. When the edge fails, the cloud revises the text and resends it. For each interaction with cloud LLM, the cloud LLM is given a prompt like ``add more details to prevent failures'' that operates on an unstructured context.

However, plain text gives the cloud no way to pinpoint what level of detail the edge actually needs. Revisions become guesswork, and the edge receives instructions it cannot reliably parse. In contrast, a tree structure offers two advantages. First, it mirrors widely-used structured formats such as Markdown and HTML, which language models already understand well~\cite{liu2026diveclaudecodedesign}. The cloud can reason about the hierarchy and the edge can descend to the depth that matches its capability, applying only the rules relevant to its current context. Second, the tree provides explicit patch locations: the cloud pinpoints exactly which node to edit, and the edge applies only the relevant rules at each level. 

% This transforms guidance from an opaque monologue into a structured, maintainable document.

\subsection{Collaborative Skill Evolution}

Atop the above two bidirectional-aware tree structures, \textsc{Co-Skill} further introduces \textit{collaborative skill evolution} to jointly evolve cloud and edge models/agents.

% It complements the trie and skill tree by jointly managing the maturity of every skill node on both cloud and edge, so that stabilized knowledge is eventually absorbed into edge parameters and pruned from the context.

\paragraph{Separated Evolution Mechanisms.}
We take separated evolution mechanisms: the cloud evolves the skill context, while the edge absorbs stabilized skills into its parameters via RL.
This division reflects the asymmetric capabilities of the two models. The cloud model is large and strong at reasoning but poorly suited for reinforcement learning due to its scale and inference cost. The edge model is small and can be fine-tuned efficiently via RL, but lacks the capacity to reason about high-level strategy. Each side operates where it excels.

\paragraph{Cloud Evolution.}
On the cloud side, skills determined through progressive refinement are not discarded after each task. Instead, they are persisted as a reusable skill library. When the edge encounters a similar task later, the cloud retrieves the matching skill directly, bypassing the progressive probing stage and substantially reducing token consumption. 

% TODO: Hot/Warm/Cold separation seems weird.
% FIX: HOT->Idel, warm->active, cold->Resident

To prevent the library from growing unbounded, we organize skills by a temperature-based scheme. New skills are marked as \emph{Idle} and discarded if success rate falls below 0.3 after at least ten uses. Skills that are frequently reused/revised stabilize to \emph{Active}. Once a skill demonstrates cross-task generalization across at least three distinct task types, it is promoted to \emph{Resident}, indicating it is stable and general. 

% \paragraph{Edge Evolution.} The progressive skill tree (\S\ref{sec:skill-tree}) first expands guidance to a sufficient depth. 
% Once the edge's success rate with this context reaches a target threshold, RL triggers to learn the policy for this skill.

% For each task, the agent samples $G$ trajectories $\{\tau^{(1)},\ldots,\tau^{(G)}\}$ from the current policy $\pi_\theta$. Each trajectory receives a binary reward $R_i \in \{0,1\}$ indicating task success, and advantages are normalized as $A_i = \frac{R_i - \mathrm{mean}(\{R_j\}_{j=1}^G)}{\mathrm{std}(\{R_j\}_{j=1}^G)}$.
% We fine-tune the edge SLM via GRPO ($G=6$) with LoRA ($r=32$, $\alpha=64$) on attention projections, using 12 training and 32 validation episodes per phase.

% Besides the normal RL process, we also modify the skill similar to Skill0~\cite{lu2026skill0incontextagenticreinforcement}. Instead of eliminating the whole skill, we progressively prune nodes from the top down upon convergence: high-level general nodes are removed first (e.g., L1 in Figure~\ref{fig:skill_tree}), then downward through L2 and L3, as general knowledge internalizes faster via RL while fine-grained steps still need skill guidance.

\paragraph{Edge Evolution.} The progressive skill tree (\S\ref{sec:skill-tree}) expands guidance to a sufficient depth, which enhances SLM's execution ability.
However, for further improvements, an optional RL module can be triggered when the edge's success rate reaches a target threshold.

Specifically, to illustrate the integration of RL into edge evolution, we design a GRPO-style RL module.
For each task, the agent samples $G$ trajectories $\{\tau^{(1)},\ldots,\tau^{(G)}\}$ from the current policy $\pi_\theta$. Each trajectory receives a binary reward $R_i \in \{0,1\}$ indicating task success, and advantages are normalized as $A_i = \frac{R_i - \mathrm{mean}(\{R_j\}_{j=1}^G)}{\mathrm{std}(\{R_j\}_{j=1}^G)}$.
The edge SLM can be fine-tuned with group size $G=6$ and LoRA ($r=32$, $\alpha=64$) on attention projections, using 12 training and 32 validation episodes per phase.

Besides the normal RL process, we also internalize the skill similar to Skill0~\cite{lu2026skill0incontextagenticreinforcement}. Instead of eliminating the whole skill, we progressively prune nodes from the top down upon convergence: high-level general nodes are removed first (e.g., L1 in Figure~\ref{fig:skill_tree}), then downward through L2 and L3, as general knowledge internalizes faster via RL while fine-grained steps still need skill guidance.
Note that Co-Skill operates without RL unless explicitly stated otherwise. 
Detailed analysis of RL can refer to~\S\ref{sec:ablation}.

\subsection{Implementation}

We use DeepSeek-V4-Pro~\cite{deepseekai2026deepseekv4highlyefficientmilliontoken} for cloud inference and Qwen3-4B-Thinking-2507~\cite{qwen3technicalreport} for edge execution on A800 GPUs. Both the prefix-merged trajectory trie and progressive skill tree are formatted as structured Markdown context. For skill evolution, we store skills in RocksDB and train the edge policy via GRPO-based RL using the \verb|verl| framework~\cite{sheng2024hybridflow}. 

\section{Evaluation}\label{sec:evaluation}

This section aims to answer the following three questions: (1) How does \textsc{Co-Skill} compare against cloud-only, local-RL, and hybrid baselines in task success rate? (\S\ref{sec:accuracy}) (2) Can \textsc{Co-Skill} reduce token usage? (\S\ref{sec:efficiency}) (3) What is the contribution of each component? (\S\ref{sec:ablation})

\begin{figure*}[t]
    \centering
    \includegraphics[width=\linewidth]{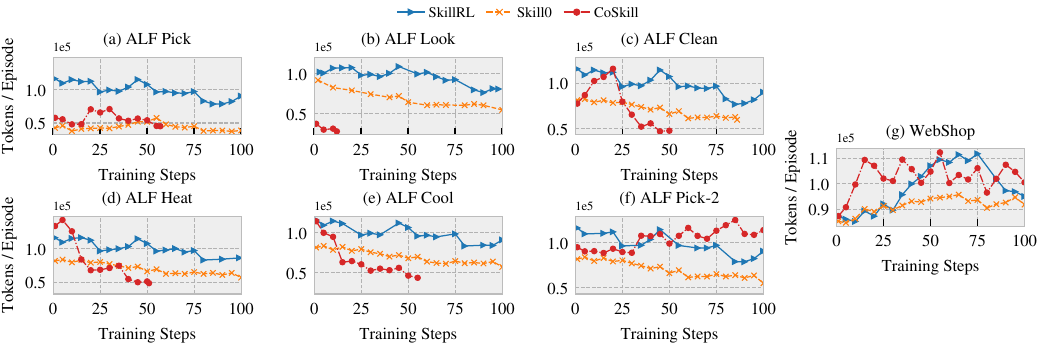}
    \caption{Average SLM+LLM token usage per episode (i.e., one complete task execution from start to finish) on ALFWorld and WebShop. Note that the WebShop workload has only one task type, so we only report one average result.}
    \label{tab:token_cost}
\end{figure*}

\subsection{Experimental Setup}

\paragraph{Competitors.}
We compare three state-of-the-art methods.
(1) \emph{Cloud-only evolution}: DeepSeek-V4-Pro~\cite{deepseekai2026deepseekv4highlyefficientmilliontoken} with carefully crafted skill sets, serving as a cloud-only upper bound without edge fine-tuning.
(2) \emph{Edge-only evolution}: Skill0~\cite{lu2026skill0incontextagenticreinforcement}, which trains the edge model with GRPO using fixed flat skills and no cloud assistance.
(3) \emph{Hybrid evolution}: SkillRL~\cite{xia2026skillrlevolvingagentsrecursive}, which dynamically updates a flat skill library with cloud feedback and fine-tunes (via RL) the edge policy with GRPO.

\paragraph{Benchmarks.}
We evaluate these evolution approaches on ALFWorld and WebShop, two widely established interactive agent benchmarks.
ALFWorld is a text-based household environment spanning six task types: pick-and-place-simple (Pick), examine-in-light (Look), clean-and-place (Clean), heat-and-place (Heat), cool-and-place (Cool), and pick-two-and-place (Pick-2). WebShop is an e-commerce benchmark where agents search, browse, and purchase products matching natural language specifications.

\paragraph{Methodology.} We adopt the default benchmark settings: each ALFWorld episode (i.e., one complete task execution from start to finish) is limited to 40 steps, and each WebShop episode to 15 steps. For each method, training stops when 100 steps are reached, or the mean success rate over the latest 10 steps exceeds 90\%, with no individual step falling below 85\%. For \textsc{Co-Skill}, we disable the RL by default, and we study how RL improves the results in~\S\ref{sec:ablation}.

\begin{figure}[t]
    \centering
    \includegraphics[width=\linewidth]{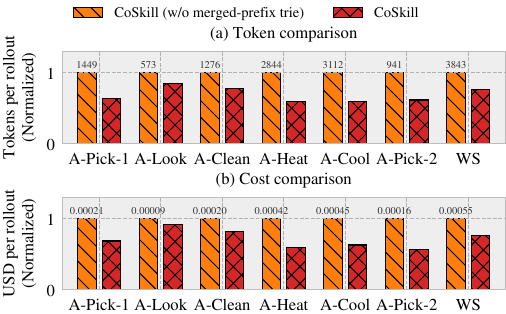}
    \caption{Prefix-merged trajectory trie efficiency analysis. Co-Skill w/o trie uploads raw trajectories. The numbers above the bar indicate the absolute values of tokens/USD.}
    \label{fig:compression}
\end{figure}

\begin{figure}[t]
    \centering
    \includegraphics[width=\linewidth]{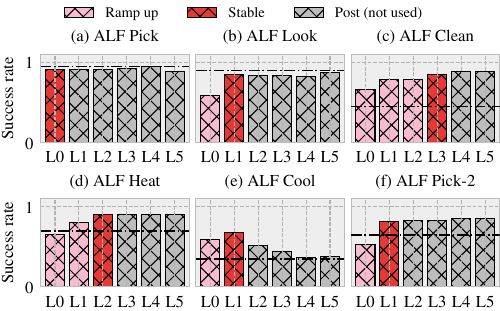}
    \caption{The effectiveness of the progressive skill tree. We disable LLM and RL to focus on only SLM execution ability. The x-axis denotes the level of the skill tree. The pink, red, and gray bars indicate unstable, stable, and overfit SLM execution capabilities. \textsc{Co-Skill} stops expanding skill tree when the success rate is stable/converges.}
    \label{fig:tree_depth}
\end{figure}

\begin{figure}[t]
    \centering
    \includegraphics[width=\linewidth]{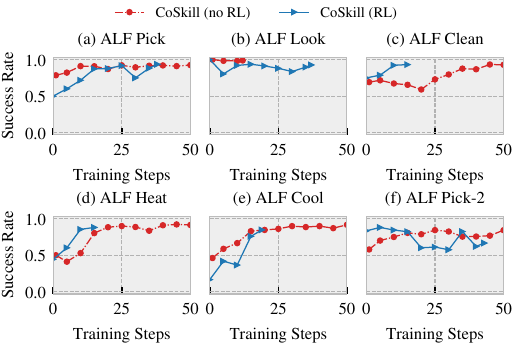}
    \caption{RL efficiency. Success rates of \textsc{Co-Skill}$_{\mathrm{noRL}}$ and \textsc{Co-Skill}$_{\mathrm{RL}}$ across the six ALFWorld task types.}
    \label{fig:rl_ablation}
\end{figure}

\subsection{Overall Accuracy}
\label{sec:accuracy}

Figure~\ref{fig:main_results} reports average task success rates on ALFWorld and WebShop. On ALFWorld, \textsc{Co-Skill} consistently surpasses all edge baselines (Skill0 and SkillRL) and approaches, and on several tasks exceeds, the cloud-only DeepSeek-V4-Pro. The advantage is most pronounced on complex multi-step tasks: on clean-and-place and cool-and-place, Cloud-Only degrades to only 45\% and 35\%, respectively, due to the lack of structured skill memory across episodes. In contrast, \textsc{Co-Skill} achieves 88.9\%--90.5\%, thanks to the collaborative communication framework that enables the cloud to iteratively refine skills and the edge to internalize them via RL. On simpler tasks such as pick-and-place and examine, all methods perform competitively, but \textsc{Co-Skill} converges more quickly.

On WebShop, all methods achieve relatively low success rates due to the tight 15-step budget, which leaves limited room for execution. Even the cloud-only DeepSeek-V4-Pro reaches only 25.0\%. Under this constrained setting, \textsc{Co-Skill} nonetheless approaches the cloud baseline and consistently outperforms all other edge methods (Skill0 and SkillRL), demonstrating that collaborative communication benefits hold even when the action space is narrow.

% Cloud-Only achieves 25.0\% success rate with an average of 49,872 total tokens per episode.
\subsection{Token Usage Efficiency}
\label{sec:efficiency}

\paragraph{SLM token usage during training.} \textsc{Co-Skill} spends $\sim$24.0\% and $\sim$7.1\% fewer tokens than SkillRL and Skill0 on average, thanks to (1) the progressive skill tree progressively matches SLM execution ability and (2) \textsc{Co-Skill} converges earlier due to the collaborative communication.

\paragraph{Cloud LLM token usage during training.} We also measure cloud LLM token usage (not shown).
Compared with SkillRL ($\sim$60 tokens per rollout), \textsc{Co-Skill} spends more tokens per rollout ($\sim$1300 tokens, but no more than USD~0.00042), because it performs more thorough and fine-grained analysis of merged trajectories and skill trees, and generates multiple rounds of skill tree. 
However, SkillRL employs aggressive truncation of the observation window, 
% discards critical contextual information, 
which compromises overall performance.

\paragraph{Overall SLM+LLM token usage.}  Despite the increased cloud tokens, as Figure~\ref{tab:token_cost} shows, the total SLM+LLM token usage is reduced by 15.6\%--41.9\% thanks to the collaborative communication adopted by \textsc{Co-Skill}. Moreover, the cloud overhead is amortizable: the skill library persists the refined skills, so subsequent executions no longer consume additional cloud LLM tokens.

\paragraph{Token usage after training.} After training stabilizes, we also measure average edge-SLM tokens per episode (not shown).
Across ALFWorld and WebShop, \textsc{Co-Skill} reduces token cost by 61.7\%--65.6\% relative to SkillRL, and by 16.2\%--41.2\% relative to Skill0.
These results indicate the effectiveness of collaborative communication in reducing per-episode token cost after convergence.

\subsection{Ablation Studies}
\label{sec:ablation}

\paragraph{Merged-prefix trajectory tree efficiency.}
Figure~\ref{fig:compression} studies the efficiency of the prefix-merged trajectory trie.
we compare the token count of the merged trie against that of naively concatenated raw trajectories from the same set of episodes under the six ALFWorld workloads and WebShop. The trie effectively reduces token consumption by merging prefixes across episodes and explicitly recording only divergence points, eliminating redundant trajectory transmission.

\paragraph{Progressive skill tree effectiveness.}
Figure~\ref{fig:tree_depth} studies the effectiveness of the progressive skill tree.
Specifically, we disable both cloud LLM and RL, then incrementally expand the tree from L0 to L5, evaluating the frozen SLM at each depth on fixed task instances. 

The results show that the optimal level varies by task type: accuracy improves to the task-specific peak, but further expansion often affects little or even degrades performance, as overly detailed instructions overwhelm the SLM.

\paragraph{Collaborative skill evolution efficiency.}
We study collaborative skill evolution from two perspectives. (1) \emph{Edge RL efficiency.} Figure~\ref{fig:rl_ablation} shows that with RL enabled, \textsc{Co-Skill} can potentially improve success rate and shorten training steps (i.e., see Figure~\ref{fig:rl_ablation}(c)), but suffer from unstable performance. (2) \emph{Cloud evolution.} When re-executing previously solved tasks, \textsc{Co-Skill} requires no additional training or skill generation---the skill library directly retrieves existing skills, eliminating cloud queries. These results demonstrate the efficiency of collaborative skill evolution: the edge absorbs what it learns, and the cloud reuses what it stores.

\section{Limitations and Future Work}

\paragraph{Emulation-based evaluation.}
Our evaluation, like most prior work~\cite{xia2026skillrlevolvingagentsrecursive,lu2026skill0incontextagenticreinforcement,ni2026trace2skilldistilltrajectorylocallessons}, relies on offline emulation in text-based environments, due to the scarcity of vision-language benchmarks with rich interaction trajectories. However, \textsc{Co-Skill}'s core idea, i.e., CCF for skill generation, refinement, and execution, should also be feasible to extend to vision language models (VLM), which we leave for future work. 

\paragraph{End-to-end online evolution.} Similar to prior works~\cite{xia2026skillrlevolvingagentsrecursive,lu2026skill0incontextagenticreinforcement,ni2026trace2skilldistilltrajectorylocallessons}, \textsc{Co-Skill} currently evaluates skill evolution offline, rather than through end-to-end online evolution where the edge agent continuously improves from live interaction feedback.
A full online deployment calls for an asynchronous RL loop: the cloud periodically updates the skill tree from accumulated trajectories while the edge concurrently executes tasks. However, it is non-trivial to achieve asynchronous RL efficiently due to the known off-policy issue~\cite{SAO, streamRL}, leaving as our future work.

\section{Conclusion}

This paper proposes \textsc{Co-Skill} for efficient edge-cloud hybrid skill evolution. The key idea is the \emph{Collaborative Communication Framework (CCF)}. Unlike prior works where the cloud and edge evolve blindly, CCF enables collaborative communication via three techniques: a prefix-merged trajectory trie, a progressive skill tree, and a collaborative skill evolution mechanism that jointly optimizes both sides. On ALFWorld and WebShop, \textsc{Co-Skill} reduces total SLM+LLM token usage by 15.6\%--41.9\% over the state-of-the-art SkillRL, while also improving task success rate.

\appendix

\bibliography{aaai2026}

\end{document}